\documentclass[prl,twocolumn,superscriptaddress,showpacs]{revtex4}

\usepackage{graphicx}
\usepackage{amsmath}
\usepackage{amsfonts}
\usepackage{amssymb}
\usepackage{bm}
\usepackage{latexsym}

\begin{document}

\title{Emergence of classicality in small number entangled systems.}

\author{Eduardo Mascarenhas}

\author{Marcelo Fran\c{c}a Santos}
 \email{msantos@fisica.ufmg.br}
 \affiliation{Dept. de F\'{\i}sica, Universidade Federal de
  Minas Gerais, Belo Horizonte,
30161-970, MG, Brazil}


\begin{abstract}
We show the transition from a fully quantized interaction to a semiclassical one in entangled
small number quantum systems using the quantum trajectories approach. In particular, we simulate the microwave Ramsey zones used in Rydberg-atom interferometry, filling in the gap between the strongly entangling Jaynes-Cummings evolution and the semiclassical rotation of the atomic internal states. We also correlate the information flowing with leaking photons to the entanglement generation between cavity field and flying atom and detail the roles played by the strong dissipation and the external driving force in preserving atomic coherence through the interaction. 
\end{abstract}

\pacs{03.65.-w, 42.50.-p, 03.67.Bg, 42.50.Lc}

\maketitle

The quantum-classical limit dilemma has been at the core of quantum theory since its beginning with arguments ranging from complementarity~\cite{Bohr} to decoherence~\cite{Zeh}. Older debates centered on the relative size of the interacting systems, i.e., a mirror was taken classically because the momentum transfer of a striking photon was insignificant compared to its inertia. Newer arguments have focused on entanglement generation, i.e., the semiclassical approximation is valid whenever a quantum system changes the physical state of another one without ever significantly entangling with it. In this framework, extreme cases such as $10^6$-photon laser fields semiclassically rotating the atomic internal states are well comprehended. But if the driving field has only one photon on average, can it still behave classically? Intuition says no. Experiments and early calculations say yes. And now quantum information theory may show how.

For example, in Rydberg-atom interferometry, two-level atoms absorb light from low energy fields of typically $<n> = 1$. In one limit, in high-Q cavities, this interaction creates lots of entanglement, causing collapses and revivals in the atomic population. In the other limit, in the microwave Ramsey zones (low-Q cavities), an equally weak coherent field behaves classically, at least for short times, preparing superpositions of atomic internal states. These two limits have been observed experimentally for decades~\cite{Haroche}. The former corresponds to the usual Jaynes-Cummings interaction~\cite{Jaynes} and the later has been recently demonstrated in~\cite{Luiz}, where the authors show that in the presence of strong dissipation and an equally strong external driving force, a very weak field can, counter-intuitively, behave in a classical way. However, most of the important questions concerning the quantum-classical transition itself remained unanswered in~\cite{Luiz}: the photon-by-photon picture of the semiclassical approximation, the specific roles of dissipation and the driving field in avoiding the inherent entanglement created by the atom-field interaction and, last but not least, the information flow with photons leaking from the low-Q cavity, and its role in the entanglement analysis.

Here, we use the quantum trajectories approach~\cite{Carmichael} to provide a deeper quantum information analysis of such a semiclassical transition. By unravelling the joint evolution of the two-level atom and the cavity field in trajectories we are able to answer all the questions raised above. The method allows us to calculate as a function of the quality factor of the cavity: how much and at what rate the reservoir reads information leaking from the system, how this information affects and entangles the atom-field quantum state, and ultimately how, at the extreme experimental conditions tested in~\cite{Luiz}, the atom and cavity fields do evolve separately. These results allow us to elucidate the roles played by the reservoir and the driving field in the semiclassical approximation, and fill in the gap between fully quantum and semiclassical behavior of low number cavity fields. Finally, we are also able to anti-correlate Environment Assisted Entanglement (EAE)~\cite{Plenio,Nha} and the semiclassical evolution, i.e., we show that the second is valid exactly when the environmental action is no longer able to create meaningful entanglement between the cavity field and the atom.

Our system consists of a cavity mode of creation operator $a^{\dagger}$, resonantly coupled to a two-level atom of raising operator $\sigma_+$. The oscillator is driven by a resonant classical field of strength $F$ and coupled to an external reservoir of decay rate $\gamma$ at zero temperature. The atom relaxation time is assumed much longer than the other time scales, which is true for fast circular Rydberg atoms inside microwave Ramsey zones. The evolution of the atom-cavity field density operator in the interaction picture is then governed by
\begin{equation}
\dot{\rho}=\frac{1}{i\hbar}[\tilde{H},\rho]-\frac{\gamma}{2}\{a^{\dagger} a,\rho\}+\gamma a\rho a^{\dagger}
\label{master}
\end{equation}
where $\tilde{H}=\hbar g(\sigma_{-}a^{\dagger}+\sigma_{+}a)+\hbar F(a^{\dagger}+a)$. 

This equation can be rewriten as 
\begin{equation}
\dot{\rho}=\frac{1}{i\hbar}[H_{eff}\rho-\rho H_{eff}^{\dagger}]+\gamma a\rho a^{\dagger},
\label{rho}
\end{equation}
which is the photon-detection unraveling of eq.(\ref{master}), where $H_{eff}=\tilde{H}-i\hbar\frac{\gamma}{2}a^{\dagger}a$ (for different unravelings, see, for example~\cite{Mabunchi}). The evolution can be analyzed for time intervals $dt$ much smaller than the typical time scales of eq. (\ref{master}) so that the density matrix at time $t + dt$ is given by
\begin{equation}
\rho(t+dt)=\sum_{j}W^{j}\rho(t)W^{j\dagger}
\end{equation}
where the set of operators $\{W^{0},W^{1},...,W^{j}\}$ are associated to the possible outcomes of the 
continuous monitoring of the reservoir. It follows that an
initial pure state $|\phi_{(t)}\rangle$ evolves, after a time $dt$, to
the (non-normalized) state
$|\phi_{(t+dt)}\rangle_{j}=W^{j}|\phi_{(t)}\rangle$ with probability
$dp_{j}=\langle\phi_{(t)}|W^{j\dagger}W^{j}|\phi_{(t)}\rangle$.
In our case, there are two possible W's: $W^{0}=I-iH_{eff}dt/\hbar$ gives the evolution for the atom-cavity system if no photon is detected in the reservoir (no-jump) and $W^{1}=\sqrt{\gamma dt}a$ dictates the evolution when a photon is detected in the reservoir (one jump). A particular realization, also called quantum trajectory, corresponds to a sequence of $W$'s applied to the initial state. For example, after a sequence of  ``n'' time intervals (corresponding to a particular trajectory $U$), the atom-cavity state will be given by 
\begin{equation}
|\phi_U(t = ndt)\rangle = \frac{W_nW_{n-1}...W_2W_1|\phi(0)\rangle}{N},
\end{equation}
where $N=\sqrt{\langle\phi(t)|\phi(t)\rangle}$ and each $W_j$ can be either $W^0$ or $W^1$ depending on the particular outcome of the $jth$ measurement of the reservoir - zero or one photon. $\rho(t)$ is recovered when summing over a large number $T$ of randomly generated  trajectories 
\begin{equation}
\rho(t)=\lim_{T \rightarrow \infty}\frac{1}{T}\sum_{U}^{T}|\phi_{U}(t)\rangle\langle\phi_{U}(t)|.
\end{equation}

The initial state considered here is $|\Psi(0)\rangle = |g\rangle|\frac{2F}{i\gamma}\rangle$, where $|g\rangle$ is the atomic internal ground state, and $|\frac{2F}{i\gamma}\rangle$ is a field coherent state with an average number of photons $\frac{4F^2}{\gamma^2}$. This is a steady state of eq.(\ref{master}) when the coupling constant $g$ is zero. In this paper, we will always consider a driving force which is equivalent to the dissipative one, i.e. $\gamma \simeq 2F$, since we are interested in low quantum number cavity fields behaving classically. State $|\Psi(0)\rangle$ is allowed to evolve in time following one particular trajectory. The routine is repeated until we have generated a sufficient large number of trajectories and then we proceed to calculate different physical quantities for the time evolved atom-field density matrix $\rho(t)$.

First, we evaluate the global purity $\delta(t)=tr\{\rho(t)^2\}$,
the atom (field) purity $\delta_{s(f)}(t)=tr\{\rho_{s(f)}^{2}(t)\}$, where
$\rho_{s(f)}$ is the atom (field) reduced density operator, the
field fidelity with its initial coherent state
$F_{c}(t)=\left\langle\frac{2F}{i\gamma}\right|\rho_{f}(t)\left|\frac{2F}{i\gamma}\right\rangle$ (Fig.1) and the Bloch evolution of the atomic state (Fig.2), for different dissipations (and equivalent driving forces). The simulation time $\Delta t=100$ is equivalent to a $2 \pi$ pulse, i.e. a complete Rabi cycle and the Bloch vector is given by 
\begin{equation}
B(x,y,z)=tr\{\rho_{s}\sigma_{x}\}\hat{x}+tr\{\rho_{s}\sigma_{z}\}\hat{z}
\label{B}
\end{equation} 
($y$ is always zero since we chose real couplings). 

In the limit $\gamma=F\simeq 0$ (pure Jaynes-Cummings evolution) the global state remains pure ($\delta(t)=1$) and $\delta_{s}(t)$ is identical to $\delta_{f}(t)$ as expected since the atom-field system is isolated and $\delta_{s(f)}(t)$ quantifies the atom-field entanglement. When $\gamma, F \gg g$ $(\gamma=200g)$ the global and atomic purity are extremely similar $\delta(t)\simeq\delta_{s}(t)$ and $\delta_{f}(t),F_{c}\simeq1$ which indicates that the atom entangles with the reservoir while the field remains in a quasi-pure state, in a
negligible statistical distance of $|\frac{2F}{i\gamma}\rangle$. This limit corresponds to the experimental conditions in \cite{Haroche}, where $g\simeq 10 kHz$, $\gamma \simeq 2 MHz$ and $|\alpha|^2\simeq 1$. In between those limits, we find that $\delta,\delta_{s},\delta_{f}$ decrease with the damping, go through a minimum at $\gamma\simeq2g$ and than increase again. This minimum indicates maximum three-partite entanglement among atom, cavity field and reservoir. It also represents the beginning of the quantum-semiclassical transition since for larger $\gamma$'s, entanglement with the reservoir will only decrease and the cavity field state rapidly converges to $|\frac{2F}{i\gamma}\rangle$. Accordingly, as shown in Fig. 2, for small cavity decay,  the atom undergoes a Bloch trajectory consistent with a Jaynes-Cummings evolution with a weak coherent field. On the other hand, for large dumping constants ($\gamma = 20g$) the atom is able to rotate from $|g\rangle$ to $(|g\rangle-|e\rangle)/\sqrt2$ almost unitarily, while for even larger decay rates ($\gamma = 200g$) the atomic evolution is almost unitary for an entire Rabi cycle. Next, we will show that this picture is explained by a single-photon description of environment-assisted entanglement.
\begin{figure}[h]
\includegraphics[height=6.5cm]{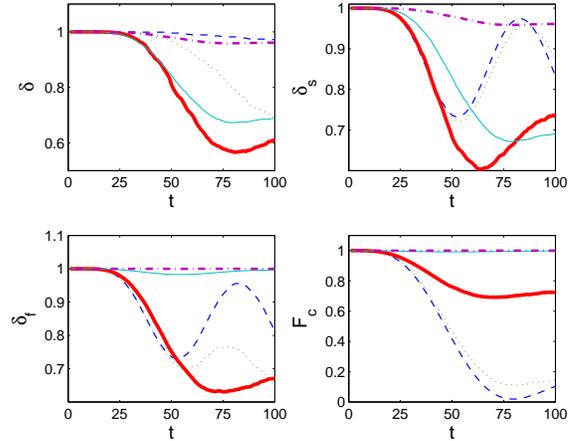}\\
  \caption{{\scriptsize Global Purity $\delta$, atom Purity $\delta_{s}$, Field Purity $\delta_{f}$ and field Fidelity $F_{c}$ for $g=10kHz$. Simulation time $\Delta t=\pi/g$. Blue dashed line for $\gamma = 0.02g$, Green doted line for $\gamma = 0.2g$, Red solid for $\gamma = 2g$, Light blue thin line $\gamma = 20g$ and Pink dashed/doted line for $\gamma = 200g$}}
  \label{pureza}
\end{figure}

\begin{figure}[h]
\includegraphics[height=5.5cm]{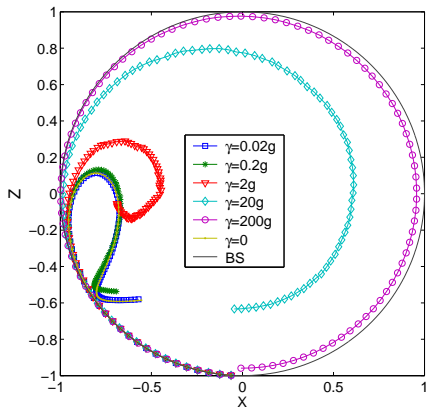}\\
  \caption{{\scriptsize Cross section $y=0$ of the Bloch Ball with the atom motion corresponding to eq. (\ref{B}): The Bloch vector $B(x,0,z)$ for different damping regimes with $\gamma=2F$. Clockwise motion starting at (0,0,-1)}}
\end{figure}

First, let us investigate some properties of the entanglement between the atom and the cavity field by calculating the average von Neumann entropy
\begin{equation}
\overline{E}=\frac{1}{T}\sum_{U}^{T}E_{U},
\label{Ebarra}
\end{equation}
with
\begin{equation}
E_{U}=-tr\{\rho_{s}^{U}\log_{2}(\rho_{s}^{U})\}
\label{EU}
\end{equation}
and $\rho_{s}^{U}=tr_{f}\{|\phi_{U}\rangle\langle\phi_{U}|\}$ as presented in Fig. 3. This average does not correspond
to the actual entanglement between these systems, it just represents the contextual entanglement
corresponding to the chosen unraveling as discussed in \cite{Nha}. However, it helps analyzing how the reservoir affects the atomic dynamics, by showing how entanglement between atom and cavity field is created on average for different trajectories.

First, note that, as shown in Fig. 1, significant entanglement between atom and cavity field only shows up after some evolution time. Second, note that when the decay rate $\gamma$ is very large, no entanglement is created between the cavity field and the atom ($\gamma=200g$). This situation corresponds to trajectories where the reservoir detects lots of photons coming from the system and yet, as shown in Figs. 1, 2 and 3, decoherence is negligible, the atom evolves unitarily and there is no entanglement created between atom and cavity field. This suggests a connection between these three effects which we will demonstrate next by analyzing single trajectories and photon-by-photon behavior of the entire system.  

\begin{figure}[h]
\begin{center}
\includegraphics[height=5.5cm]{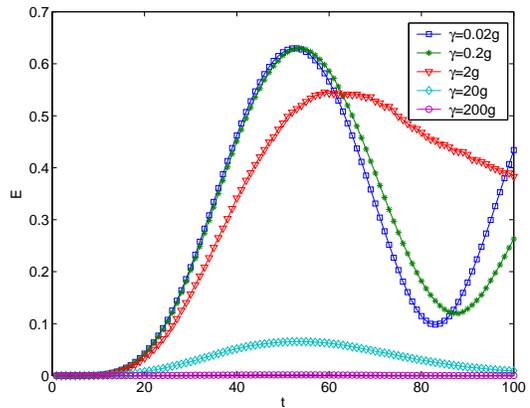}\\
\end{center}
  \caption{{\scriptsize Atom field entanglement $\overline{E}$ defined in eq.(\ref{Ebarra}) for different damping regimes with $\gamma=2F$.}}
\end{figure}

\begin{figure}[h]
\includegraphics[height=5.5cm]{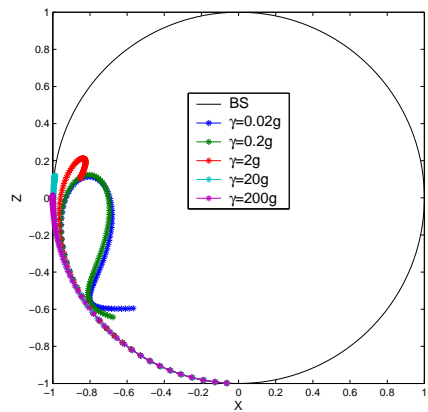}\\
  \caption{{\scriptsize The Bloch vector $B_{NJ}$ for NJT eq.(\ref{phidelta}) for different damping regimes with $\gamma=2F$.}}
  \label{nojump}
\end{figure}

To fully understand the picture, let us begin by reviewing the roles of the operators set $\{W_{j}\}$.
$W^{0}$ originates no-jump trajectories (NJT) (Fig. \ref{nojump})
\begin{equation}
|\phi(\Delta t)\rangle_{NJ}=(W^{0})^{n}|g\rangle|\alpha\rangle,
\label{phidelta}
\end{equation}
which in the continuous limit reads
\begin{equation}|\phi(t)\rangle_{NJ}=e^{-iH_{eff}t/\hbar}|g\rangle|\alpha\rangle\end{equation}
encompassing three processes: $\tilde{H}_{sf}$, $\tilde{H}_{driv}$
and $K=-i\hbar\frac{\gamma}{2}a^{\dagger}a$. When $\gamma$ is very small, this trajectory is the most probable and it yields the typical Jaynes-Cummings evolution between a two-level atom and a weak coherent field ($\gamma=0.02g$). On the other limit, for very large $\gamma 's$, the atomic motion $B_{NJ}$ for NJT is confined to the Bloch Sphere showing entanglement suppression. However, in this limit, these trajectories are very rare and we need to investigate the effect of quantum jumps on the system since typical trajectories record large numbers of them in such regimes.

We then turn to $W^{1}$ by first taking a closer look at a 
particular trajectory U (shown in Fig. 5). It is clear that quantum jumps (scattered
photons) change entanglement abruptly causing an
\emph{entanglement-leap} $\Delta E$ (similar results have been found in a different context~\cite{Carvalho2}).
This is explained by the fact that the outcome of a particular measurement, $O_{U}(t)=j(\epsilon_{t})$ is a dichotomous function of the random variable $\epsilon$ ($j=0,1$) and when $O_{U}=1$ a discontinuity takes
place in $E_{U}$. The question then is how these leaps behave as a function of the decay rate of the system and how much information is extracted by the reservoir as dissipation increases. In order to answer this question, we chose a particular time and calculated $\Delta E(\gamma)$ for a quantum jump, i.e. how much information each scattered photon carries as a function of decay rate. As shown in Fig. \ref{DeltaE}, $\Delta E(\gamma)$ decreases with $\gamma$. We also calculated the average photon counting in the reservoir 
\begin{equation}
\overline{O(t,\gamma)}=\frac{1}{T}\sum_{U}^{T}O_{U}(t,\gamma),
\end{equation}
which increases linearly with $\gamma$. The product of these two quantities $E_{L}(\gamma) = \overline{O(t,\gamma)}\Delta E(\gamma)$ represents the average information leaking to the reservoir as a function of $\gamma$.
\begin{figure}[h]
\includegraphics[height=3.8cm]{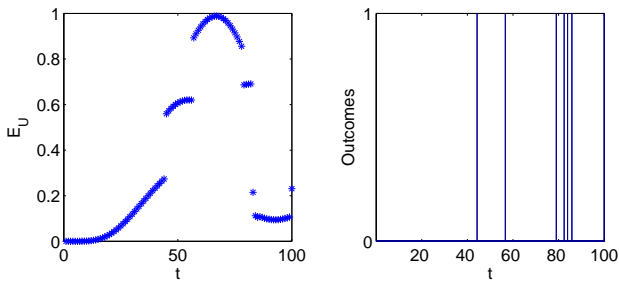}\\
  \caption{{\scriptsize On the left Entanglement eq.(\ref{EU}) for a trajectory with $\gamma=2g$ and $F=\gamma/2$, on the right the respective photon counts $O_{U}(t)$}}.
\label{leaps}
\end{figure}

When entanglement suppression ($\gamma\gg g$) is achieved a large
number of scattered photons is expected, but the leaps caused by
such photons become extremely small, i.e. detected photons do not inform on the atom-cavity system, both $W^0$ and $W^1$ become non-entangling and the mentioned classicality of the cavity field arises. 
The exact opposite happens for
low dissipation (larger leaps and infrequent scattered photons) in which case the photons
carry lots of information but they are too rare. In fact, $E_{L}$ presents the
interesting facet of EAE discussed in~\cite{Plenio, Nha}, increasing 
with the damping of the cavity, passing through a peak at
$\gamma\simeq2g$ and then decreasing asymptotically to zero. This peak corresponds to the 
red solid line in Fig.~\ref{pureza} and represents the best strategy for the reservoir to read information
stored in the atom-cavity system and, hence, the largest decoherence rate when this information is ignored. This peak also outlines the quantum-semiclassical transition from a photon-by-photon approach.  
\begin{figure}[h]
\includegraphics[height=3.8cm]{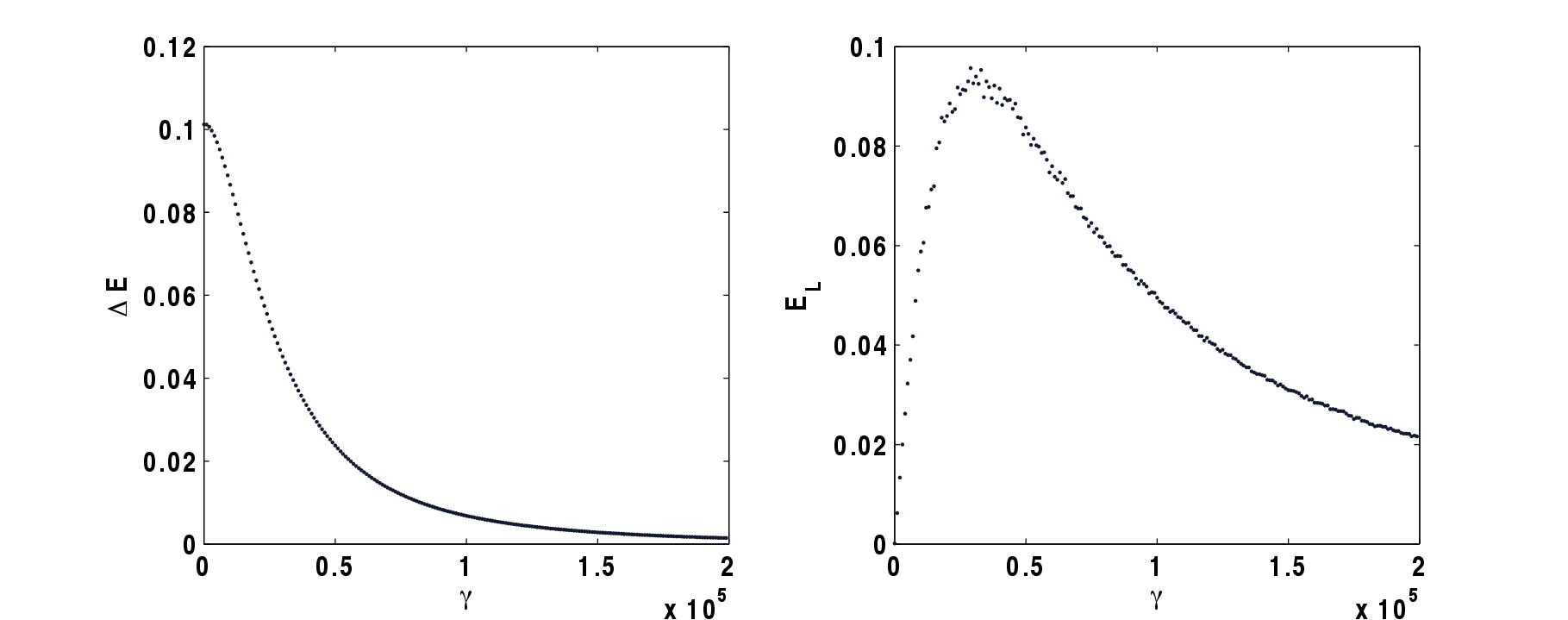}\\
  \caption{{\scriptsize  On the left entanglement leaps $\Delta E$ for different $\gamma$ (Hz) calculated for one-jump trajectories, and, on the right, a semi-quantitative estimation of the total amount of entanglement caused by leaps $E_{L}$, both for $t=\pi/4g$. $g=10KHz$}}
\label{DeltaE}
\end{figure}

We hope this paper contributes to understanding the quantum-classical limit by providing an explanation to the classical behavior of microwave Ramsey Zones from both a global and a single photon perspective. Our results allow us to attribute a measure of information carried by leaking photon and to correlate it to the entanglement generation between the low-Q cavity field and the passing atom, following an interesting EAE approach. We also show how the cavity field acts as an intermediator between the atom and the environment, remaining, in the semiclassical limit, in a quasi-pure stationary coherent state. Finally, we suggest that a similar approach may be used to describe other important experimental situations like the classical behavior of lenses, mirrors, etc. 

The authors thank CNPq and Fapemig for support and L. Davidovich, M. C. Nemes, M. O. Terra Cunha, B. Amaral and R. Rabelo for fruitful discussions.

\end{document}